\documentclass[12pt,english,floatfix,showkeys,superscriptaddress,aps,prd,preprint]{revtex4}
\usepackage[latin1]{inputenc}
\usepackage[T1]{fontenc}
\usepackage{lmodern}
\usepackage{amsmath}
\usepackage{amssymb}
\usepackage{graphicx}
\usepackage{float}
\usepackage{esint}
\usepackage{dcolumn}
\usepackage{babel}
\usepackage{csquotes}
\usepackage{color}
\usepackage{caption}
\usepackage{subcaption}
\usepackage{slashed}
\usepackage{simplewick}
\usepackage{tikz-feynman}
\usepackage[]{tikz-feynman}
\usepackage{amsmath,latexsym}
\usepackage{ragged2e}

\usepackage{cancel}

\usepackage{hyperref}
\hypersetup{
    colorlinks,
    citecolor=blue,
    filecolor=green,
    linkcolor=purple,
    urlcolor=red,
}

\usepackage{slashed}

\usepackage{hyperref}
\hypersetup{colorlinks,breaklinks,
			citecolor=[rgb]{0,0.0,1.0},
            urlcolor=[rgb]{0.0,0.0,1.0},
            linkcolor=[rgb]{0,0.5,0.9}}

\begin{document}

\title{Bhabha-like scattering in the Rarita-Schwinger model at finite temperature}

%%%%%%%%%%%%%%%%%%%%%%
\author{M. C. Ara\'{u}jo}
\email{michelangelo@fisica.ufc.br}
\affiliation{Universidade Federal do Cariri (UFCA), Av. Tenente Raimundo Rocha, \\ Cidade Universit\'{a}ria, Juazeiro do Norte, Cear\'{a}, CEP 63048-080, Brasil}
%%%%%%%%%%%%%%%%%%%%%%%%%%%%%%%%%%%%%%%%%%%%%%%%%%%%%%%%%%%%%%%%%%%%%%
%%%%%%%%%%%%%%%%%%%%%%%%%%%%%%%%%%%%%%%%%%%%%%%%%%%%%%%%%%%%%%%%%%%%%%%
\author{J. G. Lima}
\email{grimario.lima@fis.ufal.br}
\affiliation{Instituto de F\'{i}sica, Universidade Federal de Alagoas, Macei\'{o}, Alagoas 57072-900, Brazil}
% \affiliation{Departamento de F\'{i}sica Te\'{o}rica and IFIC, Centro Mixto Universidad de Valencia - CSIC. Universidad de Valencia, Burjassot-46100, Valencia, Spain.}
%%%%%%%%%%%%%%%%%%%%%%%%%%%%%%%%%%%%%%%%%%%%%%%%%%%%%%%%%%%%%%%%%%%%%%%
\author{J. Furtado}
\email{job.furtado@ufca.edu.br}
\affiliation{Universidade Federal do Cariri (UFCA), Av. Tenente Raimundo Rocha, \\ Cidade Universit\'{a}ria, Juazeiro do Norte, Cear\'{a}, CEP 63048-080, Brasil}

\author{T. Mariz}
\email{tmariz@fis.ufal.br}
\affiliation{Instituto de F\'{i}sica, Universidade Federal de Alagoas, Macei\'{o}, Alagoas 57072-900, Brazil}

\date{\today}

\begin{abstract}
In this paper, we study a Bhabha-like scattering in a massive Rarita-Schwinger model at finite temperature. The analysis is conducted at the tree level and addresses temperature effects through the thermofield dynamics formalism. We consider the usual fermion-antifermion into fermion-antifermion scattering and compute the cross-section in order to investigate the influence of the finite temperature effects.
\end{abstract}

%\pacs{04.70.-s,04.50.Kd,11.30.Cp,04.60.-m}
\keywords{Bhabha scattering, finite temperature, Rarita-Schwinger, thermofield dynamics}

\maketitle

%%%%%%%%%%%%%%%%%%%%%%%%%%%%%%%%%%%%%%%%%%%%%%%%%%%%%%%%%%%%%%%%%%%%%%%%%%%%%%%%%%%%%%
\section{Introduction}\label{intro}

Higher-spin field theories, involving fields with spin $\geq 3/2$, have been widely explored over the years in various frameworks \cite{Bengtsson:2020,Bonora:2016otz,Bonora:2017ykb,Rahman:2012thy}. These studies are particularly relevant for self-consistent models of grand unification theories \cite{Adler:2014pga}, improving the ultraviolet (UV) behavior of Einstein's gravity \cite{Campoleoni:2024ced}, and the natural appearance of such theories within the AdS/CFT correspondence \cite{Giombi:2016ejx}, among other contexts. While the free propagation of arbitrary-spin fields is generally possible, the nature of their interactions heavily depends on spin \cite{Singh:1974qz,Singh:1974rc,Fronsdal:1978rb,Fang:1978wz}. In fact, introducing interactions often gives rise to consistency challenges, such as the loss of the correct physical degrees of freedom for higher-spin particles.

Since the introduction of the Rarita-Schwinger theory for describing spin-$3/2$ fields \cite{Rarita:1941mf}, it has attracted significant phenomenological interest in various settings. Notable examples include its role in the description of gravitinos within supergravity (SUGRA) \cite{Freedman:1976py,Das:1976ct,Gates:1983nr}, scattering processes involving spin-$3/2$ particles \cite{Delgado-Acosta:2009ulg,Antoniadis:2022jjy}, the modeling of hadron resonances \cite{deJong:1992wm,Pascalutsa:1999zz,Bernard:2003xf}, and studies in Lorentz-violating frameworks \cite{Gomes:2022btc,Gomes:2023qkj}. Despite its significant physical insights and phenomenological relevance, the Rarita-Schwinger theory is known to exhibit certain limitations, which continue to be analyzed in efforts to ensure a consistent formulation \cite{Moldauer:1956zz,Johnson:1960vt,Velo:1969txo,Velo:1969bt,Aurilia:1969bg,Nath:1971wp,Adler:2015yha,Adler:2015zha}.

Beyond the description of gravitino within SUGRA context \cite{Freedman:1976py,Das:1976ct,Gates:1983nr} and the modeling of hadron resonances (e.g., the $\Delta$ resonance) \cite{deJong:1992wm,Pascalutsa:1999zz,Bernard:2003xf}, several works in the literature are addressing the general possibility of spin-3/2 leptons. In the context of composite models of quarks and leptons, an excited spin-1/2 lepton is typically regarded as the lowest-lying radial and orbital excitation. Consequently, the existence of excited spin-3/2 states is also anticipated \cite{LeiteLopes:1980mh,LeiteLopes:1980pa,LeiteLopes:1981ys}. Additionally, composite quarks and leptons within extended group structures of the standard model would naturally suggest the presence of spin-3/2 quarks and leptons \cite{Tosa:1985nn}.

From a phenomenological perspective, an excited lepton can be viewed as a heavy counterpart of the corresponding ordinary lepton, sharing the same leptonic quantum numbers. Massive spin-3/2 excited states, analogous to the heavy spin-1/2 excitations, could potentially be produced at future high-energy colliders through their effective interactions with ordinary leptons. Studies addressing the production and decay properties of charged spin-3/2 leptons have previously been conducted in \cite{LeiteLopes:1980mh,Choudhury:1984bu,Almeida:1995yp}. Also, in electron-positron collisions, excited spin-3/2 leptons can be produced in pairs, although such processes are subject to kinematic limitations. For example, in ref. \cite{Criado:2021itq} the authors investigated the interactions of higher-spin particles at high-energy colliders, while in ref. \cite{Ozdem:2024rqx} the author studied pentaquark states with spin-3/2. Additionally, the works \cite{Cakir:2007wn, Ozansoy:2016ivj} investigated the production of excited spin-3/2 particles at linear colliders and at LHeC, respectively.

As far as we are aware, the scattering processes involving spin-$3/2$ particles conducted in the literature do not take into account the finite-temperature effects and, in order to treat the scatterings properly, temperature has to be incorporated. Here, we adopt the thermofield dynamics (TFD) formalism. The topological framework of the TFD formalism enables the exploration of thermal effects across a variety of systems, including particle physics, quantum computing, and black hole physics \cite{Araujo:2022qke,Ulhoa:2023opw,Xu:2020wky}. At the heart of the TFD formalism lie two fundamental concepts: the duplication of the Hilbert space and the application of the Bogoliubov transformation \cite{Takahashi:1996zn,Umezawa:1982nv}. The resulting thermal Hilbert space (or doubled Hilbert space) facilitates the definition of a so-called thermal ground state. A crucial outcome of this construction is that, for any arbitrary operator, its statistical average can be expressed as its vacuum expectation value.

In the context of Lorentz symmetry violation, e.g., the TFD formalism has proven to be a powerful and widely adopted tool for calculating finite-temperature effects at tree level, while finite-temperature loop corrections are typically addressed using the Matsubara formalism \cite{Araujo:2023izx, Assuncao:2020cdo}. Specifically, the TFD formalism has been successfully applied to finite-temperature scattering processes in QED, encompassing both the spinorial \cite{Cabral:2024tqa, Santos:2021hed, Souza:2021uuf} and scalar sectors \cite{Araujo:2022qke, Araujo:2024hnu}. Furthermore, thermal effects within Lorentz-violating extensions of non-Abelian gauge sectors have also been analyzed using this formalism \cite{Santos:2022zfz}.  

The objective of this work is to study Bhabha scattering in a massive Rarita-Schwinger model at finite temperature. The Bhabha scattering will be investigated at tree level and the finite temperature effects will be addressed by employing the TFD formalism. We will focus on the usual fermion-antifermion in fermion-antifermion scattering, in which the cross section will be calculated to investigate the influence of the finite temperature effects.
    
This paper is organized as follows. In Sec.~\ref{RS}, we discuss the Lagrangian of the Rarita-Schwinger field coupled with the Maxwell field, including its general form, plane-wave expansion, and Feynman rules. The formalism of TFD for this model is discussed in Sec.~\ref{TFD}. In Sec.~\ref{CS}, we calculate the Bhabha-like differential cross sections at finite temperature using the TFD formalism. Finally, in Sec.~\ref{conclusion}, the main results of the paper are summarized. Throughout the paper, we have used the natural units and assumed $\eta^{\mu\nu}=diag(1,-1,-1,-1)$ as the Minkowski metric.

\section{Rarita-Schwinger Model}\label{RS}

In this section, we are interested in studying the massive Rarita-Schwinger model coupled with the Maxwell field. The Lagrangian density for this model is given by
\begin{eqnarray}\label{densmodelgeneral}
\mathcal{L} = \Bar{\psi}^{\mu}\Lambda_{\mu \nu}\psi^{\nu} -\frac{1}{4}F^{\mu\nu}F_{\mu\nu},
\end{eqnarray} 
where the operator $\Lambda_{\mu\nu}$ is written as
\begin{equation} \label{operatorL}
    \Lambda^{\mu\nu} = \frac{i}{2}\{\sigma^{\mu\nu}, (i\slashed{D} -m) \} ,
\end{equation}
with $\sigma^{\mu\nu}=\frac{i}{2}[\gamma^{\mu},\gamma^{\nu}]$ being the Dirac sigma matrix, and $D_{\mu}=\partial_{\mu}-ieA_{\mu}$ the usual covariant derivative. 

The expression (\ref{operatorL}) can be generalized by introducing an arbitrary real parameter $A$, resulting in a more general form of the operator $\Lambda^{\mu\nu}$, as follows \cite{Pilling:2004cu}:
\begin{eqnarray}\label{generalcaseL}
 \Lambda^{\mu\nu}&=&(i\slashed{D}-m)\eta^{\mu\nu}
+A(\gamma^{\mu}iD^{\nu}+\gamma^{\nu}iD^{\mu})+\frac{1}{2}(3A^2+2A+1)\gamma^{\mu}i\slashed{D}\gamma^{\nu}\nonumber\\ 
&&+ m(3A^2+3A+1)\gamma^{\mu}\gamma^{\nu}.
\end{eqnarray}
For the specific case where $A=-1$, we have
\begin{eqnarray}
    \Lambda_{\mu \nu} = (i \slashed{D}-m)\eta_{\mu \nu}-(\gamma_{\mu}iD_{\nu}+\gamma_{\nu}iD_{\mu})+\gamma_{\mu}i\slashed{D}\gamma_{\nu}+m\gamma_{\mu}\gamma_{\nu},
\end{eqnarray}
which represents the more commonly used definition of $\Lambda^{\mu\nu}$, as also stated in Eq.~(\ref{operatorL}). It is worth noticing that the original form of the free Lagrangian proposed in \cite{Rarita:1941mf} takes the value $A=-1/3$.

The plane-wave expansion for the Rarita-Schwinger field can be written as
\begin{eqnarray}\label{psitempzero1}
   \psi_{\mu}(x) &=& \int \frac{d^3k}{(2\pi)^3}\frac{1}{\sqrt{2E(\mathbf{k})}}\sum_s \left[ a^s(\mathbf{k})\mathbf{u}_{\mu}^s(k)e^{-ik\cdot x}+b^{s\, \dagger}(\mathbf{k})\mathbf{v}_{\mu}^s(k)e^{ik\cdot x} \right],
 \end{eqnarray}
where $a^s$ and $b^{s\, \dagger}$ are annihilation and creation operators, respectively, and $\mathbf{u}_{\mu}^s(k)$ and $\mathbf{v}_{\mu}^s(k)$ represent the Rarita-Schwinger spinor \cite{Das:1976ct}.

Let us now consider the Feynman rules for the scattering process we are interested in investigating. For the fermion-photon vertex and the photon propagator, respectively, we write
\begin{eqnarray}\label{firstvertex}
   \begin{tikzpicture}[baseline=(b3.base)]
 \begin{feynman}
    \vertex (a1);
    \vertex[right=1.2 of a1] (a2);
    \vertex[right=1.2 of a2] (a3);
    \vertex[below=1.2 of a1] (b1);
    \vertex[right=1.2 of b1] (b2);
    \vertex[right=1.6 of b2] (b3) {\(\, \, \, \alpha \mu \nu\)};
    \vertex[below=1.2 of b1] (c1);
    \vertex[right=1.2 of c1] (c2);
    \vertex[right=1.2 of c2] (c3);
    \diagram* {
      (a1) -- [fermion] (b2),
      (b2) -- [photon,] (b3),
      (c1) -- [anti fermion] (b2),
      };
  \end{feynman}
 \end{tikzpicture} = ie(\gamma_{\alpha}\eta_{\mu \nu}-\gamma_{\mu}\eta_{\alpha \nu}-\gamma_{\nu}\eta_{\alpha \mu}+\gamma_{\mu}\gamma_{\alpha}\gamma_{\nu}) = \Sigma_{\alpha \mu \nu}, 
\end{eqnarray}

\begin{eqnarray}\label{photonprop}
 \begin{tikzpicture}[baseline=(b2.base)]
 \begin{feynman}
    \vertex (b1) {\(\mu\)};
    \vertex [right= 2.6cm of b1] (b2) {\(\nu\)};
    
    \diagram* {
      (b1) -- [photon,  momentum=\(q\)] (b2), 
    };
  \end{feynman}
 \end{tikzpicture} = \frac{-i\eta_{\mu \nu}}{q^2}.
\end{eqnarray}
Note that we have added the gauge-fixing term $-\frac{1}{2\alpha}(\partial_\mu A^\mu)$, with $\alpha=1$, so that we can invert the photon propagator.

\section{thermofield dynamics formalism}\label{TFD}

The introduction of temperature in our analysis will be achieved through the thermofield dynamics (TFD) formalism. The main feature of this approach is the construction of a thermal vacuum state $|\, 0(\beta)\, \rangle$ ($\beta =1/T$, being $T$ the temperature) via the duplication of the degrees of freedom of the system. Due to this doubled structure, characterized by $\widehat{\mathcal{S}} = \mathcal{S}\otimes \widetilde{\mathcal{S}}$, associated with each operator $\mathcal{O}$ in the $\mathcal{S}$-space, where physical quantities are indeed defined, there exists a corresponding operator $\widehat{\mathcal{O}}$ in $\widehat{\mathcal{S}}$ defined by
\begin{eqnarray}
\widehat{\mathcal{O}} \equiv \mathcal{O} - \widetilde{\mathcal{O}}.
\end{eqnarray}
Here, $\widetilde{\mathcal{O}}$ is an operator that acts in the $\widetilde{\mathcal{S}}$-space, which is considered a copy of the original space $\mathcal{S}$. Tilde and non-tilde operators are connected by the following conjugation rules:
\begin{eqnarray}\label{rct1}
\widetilde{(\mathcal{O}_i\mathcal{O}_j)} &=& \widetilde{\mathcal{O}}_i\widetilde{\mathcal{O}}_j ,\\
\widetilde{(a\mathcal{O}_i+b\mathcal{O}_j)} &=& a^{\ast}\widetilde{\mathcal{O}}_i+b^{\ast}\widetilde{\mathcal{O}}_j,\\
(\widetilde{\mathcal{O}}_i)^{\dagger} &=& \widetilde{(\mathcal{O}_i^{\dagger})},\\
\label{rct4}
\widetilde{(\widetilde{\mathcal{O}}_i)} &=& \kappa \mathcal{O}_i,
\end{eqnarray} where $\kappa=1\, (-1)$ for bosons (fermions). As a consequence, thermal operators can be defined in terms of tilde and non-tilde operators through the so-called Bogoliubov transformations. In a fermionic theory, for instance, one can write
 \begin{eqnarray}\label{opefermionbetazero1}
  a(\mathbf{k},\beta) &=& u_F(\mathbf{k},\beta)a(\mathbf{k}) - v_F(\mathbf{k},\beta)\widetilde{a}^{\dagger}(\mathbf{k}); \\ \label{opefermionbetazero2}
a^{\dagger}(\mathbf{k},\beta) &=& u_F(\mathbf{k},\beta)a^{\dagger}(\mathbf{k}) - v_F(\mathbf{k},\beta)\widetilde{a}(\mathbf{k}); \\ \label{opefermionbetazero3}
\widetilde{a}(\mathbf{k},\beta) &=& u_F(\mathbf{k},\beta)\widetilde{a}(\mathbf{k}) + v_F(\mathbf{k},\beta)a^{\dagger}(\mathbf{k});\\ \label{opefermionbetazero4}
\widetilde{a}^{\dagger}(\mathbf{k},\beta) &=& u_F(\mathbf{k},\beta)\widetilde{a}^{\dagger}(\mathbf{k}) + v_F(\mathbf{k},\beta)a(\mathbf{k}),
 \end{eqnarray} 
where
\begin{eqnarray}\label{defufermioncostheta1}
  u_F(\mathbf{k},\beta) &=& e^{\beta\, E(\mathbf{k})/2}\, v_F(\mathbf{k},\beta), 
 \end{eqnarray}
 and
\begin{eqnarray}
  \label{defufermioncostheta2}
  v_F(\mathbf{k},\beta) &=& \frac{1}{\sqrt{1+e^{\beta\, E(\mathbf{k})}}}.
\end{eqnarray} 
For a bosonic theory, in turn, 
\begin{eqnarray}
\label{oprtermoperzero1}
a(\mathbf{k},\beta) &=& u_B(\mathbf{k},\beta)a(\mathbf{k}) - v_B(\mathbf{k},\beta)\widetilde{a}^{\dagger}(\mathbf{k}) ;\\ \label{oprtermoperzero2}
a^{\dagger}(\mathbf{k},\beta) &=& u_B(\mathbf{k},\beta)a^{\dagger}(\mathbf{k}) - v_B(\mathbf{k},\beta)\widetilde{a}(\mathbf{k}); \\ \label{oprtermoperzero3}
\widetilde{a}(\mathbf{k},\beta) &=& u_B(\mathbf{k},\beta)\widetilde{a}(\mathbf{k}) - v_B(\mathbf{k},\beta)a^{\dagger}(\mathbf{k});\\ \label{oprtermoperzero4}
\widetilde{a}^{\dagger}(\mathbf{k},\beta) &=& u_B(\mathbf{k},\beta)\widetilde{a}^{\dagger}(\mathbf{k}) - v_B(\mathbf{k},\beta)a(\mathbf{k}),
\end{eqnarray}
where
\begin{eqnarray}\label{funcaoubossoncosh1}
u_{B}(\mathbf{k},\beta) &=& e^{\beta \, E(\mathbf{k})/2}v_B(\mathbf{k},\beta),
\end{eqnarray} and
\begin{eqnarray}
    \label{funcaoubossoncosh2}
v_{B}(\mathbf{k},\beta) &=& \frac{1}{\sqrt{e^{\beta\,  E(\mathbf{k})}-1}}.
\end{eqnarray}
In the above expressions, $E(\mathbf{k})$  is the energy of oscillation associated with the mode $\mathbf{k}$, and $a\, (\widetilde{a})$ and $a^{\dagger}\, (\widetilde{a}^{\dagger})$ stand for annihilation and creation operators in the space $\mathcal{S}$ ($\widetilde{\mathcal{S}}$), respectivally. Furthermore, it can be shown that fermionic thermal operators satisfy \begin{eqnarray}
\label{algfermion1}
   \left\{ a^{s}(\mathbf{k},\beta),a^{r\, \dagger}(\mathbf{k'},\beta)\right\} &=& (2\pi)^3\delta^3(\mathbf{k}-\mathbf{k'})\delta^{sr};\\
\label{algfermion2} \left\{ \widetilde{a}^{\, s}(\mathbf{k},\beta),\widetilde{a}^{\, r\, \dagger}(\mathbf{k'},\beta)\right\} &=& (2\pi)^3\delta^3(\mathbf{k}-\mathbf{k'})\delta^{sr},
 \end{eqnarray}
 with all the other anticommutation relations being zero. Note that we have made the dependence on spin explicit through the superscripts $s$ and $r$. For bosons, the algebra in Eqs. \eqref{algfermion1} and \eqref{algfermion2} should be replaced by 
\begin{eqnarray}\label{comutanicriaexpress1}
    \left[ a^{\lambda}(\mathbf{k},\beta),a^{\lambda' \, \dagger}(\mathbf{k'},\beta)\right] &=& (2\pi)^3\delta^3(\mathbf{k}-\mathbf{k'})\delta^{\lambda \, \lambda'}; \\ \label{comutanicriaexpress2}
    \left[ \widetilde{a}^{\, \lambda}(\mathbf{k},\beta),\widetilde{a}^{\, \lambda'\, \dagger}(\mathbf{k'},\beta)\right] &=& (2\pi)^3\delta^3(\mathbf{k}-\mathbf{k'})\delta^{\lambda \lambda'},
\end{eqnarray}
with all the other commutation relations being zero as well. Here, $\lambda$ accounts for some physically relevant property of interest. In the case of the electromagnetic field, for example, $\lambda$ could represent the photon polarization state.

Regarding the Feynman rules discussed in the previous section, the TFD formalism introduces only modifications to the propagator. Defined in the $\mathcal{S}$-space as 
\begin{eqnarray}\label{propfotontermaldef1}
\langle\, 0(\beta)\, |\mathcal{T}\left[ A_{\mu}(x)A_{\nu}(y) \right]|\, 0(\beta)\, \rangle ,
\end{eqnarray}
with
\begin{eqnarray}\label{campodefotonszertotemp}
 A_{\mu}(x) = \sum_{\lambda=0}^{3}\int \frac{d^3k}{(2\pi)^3}\, \frac{1}{\sqrt{2E(\mathbf{k})}}\left[ a^{\lambda}(\mathbf{k})e^{-ik\cdot x} + a^{\lambda\, \dagger}(\mathbf{k})e^{ik\cdot x} \right]\epsilon^{\lambda}_{\mu}(k),
 \end{eqnarray}
representing the usual expansion for the gauge field and $\epsilon^\lambda_{\mu}(k)$ the respective polarization vector, by means of the standard canonical procedure, and through relations \eqref{oprtermoperzero1}-\eqref{funcaoubossoncosh2}, \eqref{comutanicriaexpress1} and \eqref{comutanicriaexpress2}, we can write \begin{eqnarray}\label{thermalphotonprop}
 \begin{tikzpicture}[baseline=(b2.base)]
 \begin{feynman}
    \vertex (b1) {\(\mu\)};
    \vertex [right= 2.6cm of b1] (b2) {\(\nu\)};
    
    \diagram* {
      (b1) -- [photon,  momentum=\(q\)] (b2), 
    };
  \end{feynman}
 \end{tikzpicture} = \frac{-i\eta_{\mu \nu}}{q^2} - 2\pi\, \eta_{\mu \nu}v_B^2(\mathbf{q},\beta)\delta(q^2).
\end{eqnarray}
On the same line, it can be shown that the propagator in the $\widetilde{\mathcal{S}}$-space, defined as in Eq. \eqref{propfotontermaldef1} by replacing $A_{\mu}$ with $\widetilde{A}_{\mu}$, can be derived simply by acting directly with the conjugation rules \eqref{rct1}-\eqref{rct4} on Eq. \eqref{thermalphotonprop}. The same applies to the interaction vertex, which in the $\mathcal{S}$-space take the form presented in Eq. \eqref{firstvertex}. 

\section{Bhabha-like differential cross section}\label{CS}

In this section, we will calculate the differential cross-section at finite temperature using the TFD formalism for the Bhabha-like scattering involving spin-3/2 particles. When expressed in the center-of-mass reference frame, the temperature-dependent differential cross-section in the case where all involved particles have identical masses takes the form
\begin{eqnarray}\label{secdifbetacmsomaspinmhatcmbetaquadrado}
      \left( \frac{d \sigma }{d\Omega}   \right)_{\beta} = \frac{1}{64\, \pi^2 \, E_{cm}^2}\, \frac{1}{16}\sum_{spin}|\widehat{\mathcal{M}}(\beta)|^2,
\end{eqnarray}
where
\begin{eqnarray}\label{mhatmbetamenosmtilbetadef}
 \widehat{\mathcal{M}}(\beta) = \mathcal{M}(\beta) - \widetilde{\mathcal{M}}(\beta) 
\end{eqnarray}
with
\begin{equation}\label{matrizmbetaoriginal}
   \mathcal{M}(\beta) \equiv \sum_{n=0}^{\infty}\frac{(-i)^n}{n!}\int\, d^4z_1\cdots d^4z_n\, {}_{\beta}\langle \, f\, |\mathcal{T}\left[ {\mathcal{H}}_{int}(z_1)\cdots {\mathcal{H}}_{int}(z_n) \right]|\, i\, \rangle_{\beta}
\end{equation} 
and
\begin{equation}\label{matrizbetatil}
   \widetilde{\mathcal{M}}(\beta) \equiv \sum_{n=0}^{\infty}\frac{\widetilde{(-i)^n}}{n!}\int\, d^4z_1\cdots d^4z_n\, {}_{\beta}\langle \, f\, |\mathcal{T}\left[ \widetilde{\mathcal{H}}_{int}(z_1)\cdots \widetilde{\mathcal{H}}_{int}(z_n) \right]|\, i\, \rangle_{\beta},
\end{equation}
defining the temperature-dependent transition amplitudes for the scattering of interest. Here, $E_{cm}$ is the center-of-mass energy, ${\mathcal{H}}_{int}$ ($\widetilde{\mathcal{H}}_{int}$) is the interaction Hamiltonian density in the space ${\mathcal{S}}$ ($\widetilde{\mathcal{S}}$), and $\mathcal{T}$ is the time-ordering operator. For the scattering process of interest, the initial and final particle states are respectively defined by 
\begin{eqnarray}
  |\, i\, \rangle_{\beta} &\equiv& \sqrt{2E(\mathbf{p})\, 2E(\mathbf{p}')}\, a^{s\, \dagger}(\mathbf{p},\beta)\, b^{s'\, \dagger}(\mathbf{p}',\beta)\, |\, 0(\beta)\, \rangle
\end{eqnarray}
and 
\begin{eqnarray}
    {}_{\beta}\langle \, f\, | &\equiv& \langle \, 0(\beta)\, | \, b^{r'}(\mathbf{k'},\beta)\, a^{r}(\mathbf{k},\beta) \sqrt{2E(\mathbf{k})\,  2E(\mathbf{k}')}.
\end{eqnarray} In this way, to ensure an appropriate treatment of Eq. \eqref{mhatmbetamenosmtilbetadef}, we must employ the Bogoliubov transformations in Eqs. \eqref{opefermionbetazero1}-\eqref{opefermionbetazero4} and \eqref{oprtermoperzero1}-\eqref{oprtermoperzero4} to the  amplitudes in Eqs. \eqref{matrizmbetaoriginal} and \eqref{matrizbetatil}. In short, the procedure to be followed is quite standard in quantum field theory. First, we draw all the Feynman diagrams corresponding to the scattering of interest. Then, we apply the rules presented in the previous section to write the scattering matrix defined in Eq. \eqref{mhatmbetamenosmtilbetadef}. Note that the external lines of the diagrams are typically identified through the action of the fields on the initial and final particle states. Thus, for example, an external line of a spin-$3/2$ fermion at the beginning of the process is represented in the usual way in the $\mathcal{S}$-space by the contraction 
\begin{eqnarray}
    \contraction{}{\psi}{_{\mu}(z)|\mathbf{p}}{,} \psi_{\mu}(z)|\mathbf{p},s\rangle_{\beta} &=& u_F(\mathbf{p},\beta)\mathbf{u}_{\mu}^s(p),
\end{eqnarray}
while in the $\widetilde{\mathcal{S}}$-space, it is represented by
\begin{eqnarray}
     \contraction{}{\tilde{\bar{\psi}}}{_{\mu}(z)|\mathbf{p}}{,} \tilde{\bar{\psi}}_{\mu}(z)|\mathbf{p},s\rangle_{\beta} &=& -v_F(\mathbf{p},\beta)\bar{\mathbf{u}}_{\mu}^{\ast \, s}(p),
\end{eqnarray} where 
\begin{eqnarray}
    |\, \mathbf{p},s \, \rangle_{\beta} = \sqrt{2E(\mathbf{p})}\, a^{s\, \dagger}(\mathbf{p},\beta)\, |\, 0(\beta)\, \rangle, 
\end{eqnarray} for short.

\begin{figure}[!]
    \centering
    \begin{minipage}{0.45\textwidth}
        \includegraphics[width=\linewidth]{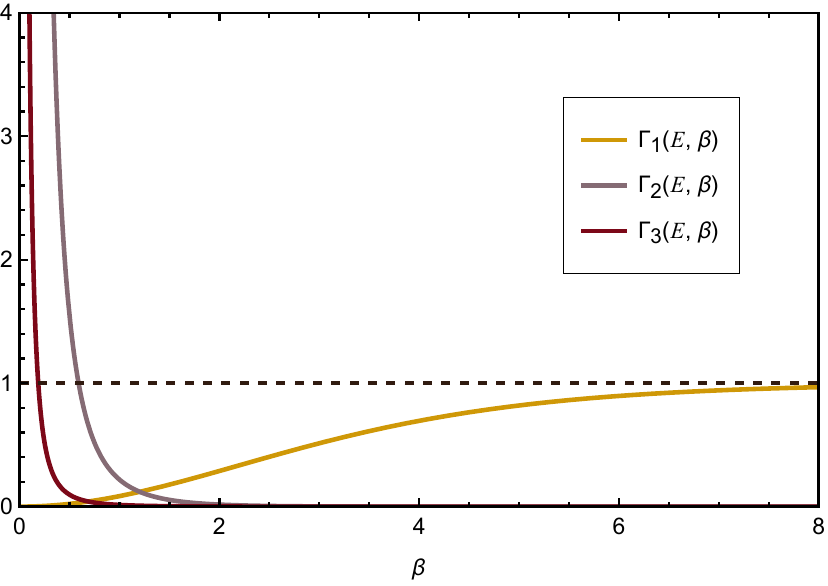}
        \captionsetup{width=\linewidth}
        \caption{\justifying Dependence of the functions $\Gamma_i$ ($i=1,2,3$) on $\beta$. We have used $E=0.6$ for this plot.}
        \label{graf1}
    \end{minipage}
    \hfill
    \begin{minipage}{0.47\textwidth}
        \includegraphics[width=\linewidth]{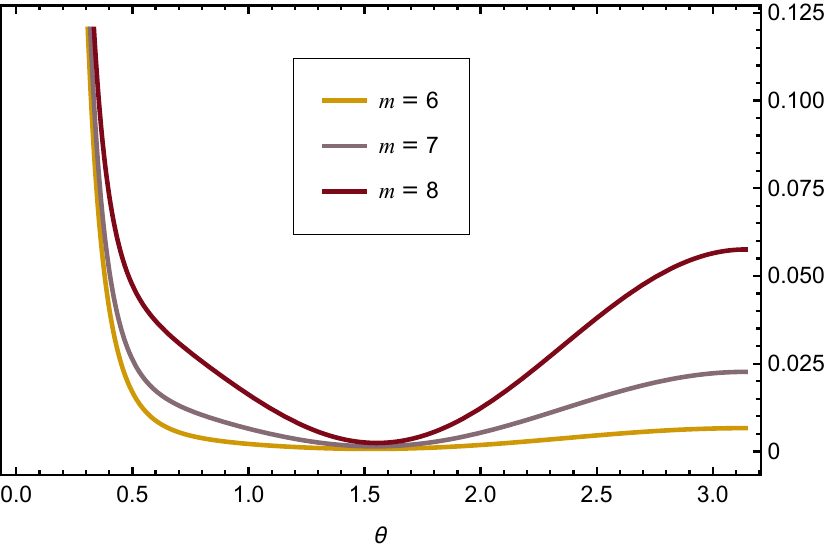}
        \captionsetup{width=\linewidth}
        \caption{\justifying Angular dependence of the differential cross-section at zero temperature for $E<m$. We have used $E=2$ and $e=1$.}
        \label{graf2}
    \end{minipage}
\end{figure}

Now, the Feynman diagrams common to both tilde and non-tilde spaces for the Bhabha-like scattering in the Rarita-Schwinger model (coupled with Maxwell) at tree level are 
\begin{eqnarray}\label{graffeyntreelschannel}
\begin{tikzpicture}[baseline=(b3.base)]
 \begin{feynman}
    \vertex (a1);
    \vertex [right=1.3 of a1] (a2);
    \vertex [right=1.3 of a2] (a3);
    \vertex [right=1.3 of a3] (a4);
    \vertex [below=1.3 of a1] (b1);
    \vertex [right=1.3 of b1] (b2);
    \vertex [right=1.3 of b2] (b3);
    \vertex [right=1.3 of b3] (b4);
    \vertex [below=1.3 of b1] (c1);
    \vertex [right=1.3 of c1] (c2);
    \vertex [right=1.3 of c2] (c3);
    \vertex [right=1.3 of c3] (c4);

    \diagram* {
      (a1) -- [anti fermion, edge label'=\(p'\)] (b2),
      (b2) -- [photon, edge label=\(q\)] (b3), 
      (b3) -- [anti fermion, edge label'=\(k'\)] (a4),
      (c1) -- [fermion, edge label=\(p\)] (b2),
      (b3) -- [fermion, edge label=\(k\)] (c4),
    };
  \end{feynman}
 \end{tikzpicture}\hspace{2cm}
 \begin{tikzpicture}[baseline=(b5.base)]
 \begin{feynman}
    \vertex (a1);
    \vertex [below=1.0 of a1] (a2);
    \vertex [below=1.0 of a2] (a3);
    \vertex [below=1.0 of a3] (a4);
    \vertex [right=1.4 of a1] (b1);
    \vertex [below=1.0 of b1] (b2);
    \vertex [below=1.0 of b2] (b3);
    \vertex [below=1.0 of b3] (b4);
    \vertex[below=0.5 of b2] (b5);
    \vertex [right=1.4 of b1] (c1);
    \vertex [below=1.0 of c1] (c2);
    \vertex [below=1.0 of c2] (c3);
    \vertex [below=1.0 of c3] (c4);

    \diagram* {
      (a1) -- [anti fermion, edge label'=\(p'\)] (b2),
      (b2) -- [photon, edge label=\(q'\)] (b3), 
      (b3) -- [anti fermion, edge label'=\(p\)] (a4),
      (c1) -- [fermion, edge label=\(k'\)] (b2),
      (b3) -- [fermion, edge label=\(k\)] (c4),
    };
  \end{feynman}
 \end{tikzpicture}.
\end{eqnarray}
On the left, we have the diagram corresponding to the s-channel, and on the right, the one corresponding to the t-channel. Note that the thermal scattering matrix in Eq. (32) must take into account these two contributions and, therefore, must be written as 
\begin{eqnarray}
    \widehat{\mathcal{M}}(\beta) = \widehat{\mathcal{M}}_s(\beta)-\widehat{\mathcal{M}}_t(\beta), 
\end{eqnarray}
where the negative sign is the overall sign associated with t-channel diagrams. Using the Feynman rules, we can then write
\begin{eqnarray}\label{mhatbetatotal301}
    \widehat{\mathcal{M}}_s(\beta) &=& \frac{1}{q^2} \left(\prod v_F\right)\left(1-\prod \frac{u_F}{v_F}\right)\left\lbrace 1+2 \pi i\,  v_B^2(\mathbf{q},\beta)\, q^2\, \delta(q^2)\left[ \frac{1+\prod \frac{u_F}{v_F}}{1-\prod \frac{u_F}{v_F}} \right]\right\rbrace  \nonumber\\
    &\times& \big[ \bar{\mathbf{v}}_{\alpha}^{s'}(p')\, \Sigma^{\mu \alpha \beta}\, \mathbf{u}_{\beta}^{s}(p) \big]\big[ \bar{\mathbf{u}}_{\gamma}^{r}(k)\, \Sigma_{\mu}{}^{\gamma \sigma}\, \mathbf{v}_{\sigma}^{r'}(k') \big]
\end{eqnarray}
and
\begin{eqnarray}\label{mhatbetatotal302}
    \widehat{\mathcal{M}}_t(\beta) &=& \frac{1}{q'^2} \left(\prod v_F\right)\left(1-\prod \frac{u_F}{v_F}\right)\left\lbrace 1+2 \pi i\,  v_B^2(\mathbf{q'},\beta)\, q'^2\, \delta(q'^2)\left[ \frac{1+\prod \frac{u_F}{v_F}}{1-\prod \frac{u_F}{v_F}} \right]\right\rbrace  \nonumber\\
    &\times& \big[ \bar{\mathbf{u}}_{\alpha}^{r}(k)\, \Sigma^{\mu \alpha \beta}\, \mathbf{u}_{\beta}^{s}(p) \big]\big[ \bar{\mathbf{v}}_{\gamma}^{s'}(p')\, \Sigma_{\mu}{}^{\gamma \sigma}\, \mathbf{v}_{\sigma}^{r'}(k') \big]
\end{eqnarray}
where
\begin{eqnarray}
    \prod v_F \equiv v_F(\mathbf{p},\beta)\, v_F(\mathbf{p'},\beta)\, v_F(\mathbf{k},\beta)\, v_F(\mathbf{k'},\beta), 
\end{eqnarray} and
\begin{eqnarray}
    \prod \frac{u_F}{v_F} \equiv \frac{u_F(\mathbf{p},\beta)\, u_F(\mathbf{p'},\beta)\, u_F(\mathbf{k},\beta)\, u_F(\mathbf{k'},\beta)}{v_F(\mathbf{p},\beta)\, v_F(\mathbf{p'},\beta)\, v_F(\mathbf{k},\beta)\, v_F(\mathbf{k'},\beta)}.
\end{eqnarray}
Once we have Eqs. \eqref{mhatbetatotal301} and \eqref{mhatbetatotal302} at hand, we can directly calculate 
\begin{eqnarray}
    |\widehat{\mathcal{M}}(\beta)|^2 = |\widehat{\mathcal{M}}_s(\beta)|^2 - \widehat{\mathcal{M}}_s(\beta)\widehat{\mathcal{M}}^{\ast}_t(\beta)-\widehat{\mathcal{M}}_t(\beta)\widehat{\mathcal{M}}^{\ast}_s(\beta)+|\widehat{\mathcal{M}}_t(\beta)|^2
\end{eqnarray} and evaluate the result from the center-of-mass reference frame of the system, where 
\begin{eqnarray}\label{setcentermassframe33}
p=(E,|\mathbf{p}|\, \hat{\mathbf{z}}); &\hspace{1cm}& k=(E,\mathbf{k}); \nonumber\\
p'=(E,-|\mathbf{p}|\, \hat{\mathbf{z}}); &\hspace{1cm}&  k'=(E,-\mathbf{k}); 
\end{eqnarray} with
\begin{eqnarray}
 \mathbf{k} = |\mathbf{k}|(\sin \theta \, \cos \phi \, \hat{\mathbf{x}}+\sin \theta \, \sin \phi \, \hat{\mathbf{y}}+ \cos \theta \, \hat{\mathbf{z}});
\end{eqnarray}
\begin{eqnarray}
   |\mathbf{p}|=|\mathbf{k}|=\sqrt{E^2-m^2}, 
\end{eqnarray}
and $E_{cm}=2E$. When performing the calculations above, it is also necessary to use projectors
\begin{eqnarray}
   \sum_s \mathbf{u}_{\mu}^s(p)\bar{\mathbf{u}}_{\nu}^s(p) &=& (\slashed{p} +m)\bigg[ \eta_{\mu \nu}-\frac{1}{3}\gamma_\mu \gamma_{\nu}-\frac{1}{3m}(\gamma_{\mu}p_{\nu}-\gamma_{\nu}p_{\mu})-\frac{2}{3}\frac{p_{\mu}p_{\nu}}{m^2} \bigg]
\end{eqnarray}
and
\begin{eqnarray}
   \sum_s \mathbf{v}_{\mu}^s(p)\bar{\mathbf{v}}_{\nu}^s(p) &=& (\slashed{p} - m)\bigg[ \eta_{\mu \nu}-\frac{1}{3}\gamma_\mu \gamma_{\nu}+\frac{1}{3m}(\gamma_{\mu}p_{\nu}-\gamma_{\nu}p_{\mu})-\frac{2}{3}\frac{p_{\mu}p_{\nu}}{m^2} \bigg],
\end{eqnarray}
in order to properly handle the resulting spinor products. Finally, when evaluated in the center-of-mass reference frame, the temperature-dependent differential cross-section for the Bhabha-like scattering within the Rarita-Schwinger model can be expressed as
\begin{eqnarray}\label{secdiftermalfinal1111}
   \left( \frac{d\sigma}{d\Omega} \right)_{\beta} &=&  \frac{e^4}{41472\,  \pi ^2\,  E^6\,  m^8} \Big[ \Pi_1(E,m,\theta)\, \Gamma_1(E,\beta)\nonumber\\
   &+& 4 \pi^2 E^4 \Pi_2(E,m,\theta)\, \Gamma_2(E,\beta)\, \delta \left((E^2-m^2) (\cos (\theta )-1)\right)^2\nonumber\\
   &+& 2 \pi ^2 E^4 \Pi_3(E,m,\theta)\, \Gamma_2(E,\beta)\, \delta \left(E^2\right) \delta \left(\left(E^2-m^2\right) (\cos (\theta )-1)\right)\nonumber\\
   &+& 32 \pi^2 E^4 \Pi_4(E,m,\theta)\, \Gamma_3(E,\beta)\, \delta \left(\text{En}^2\right)^2 \Big],
\end{eqnarray}
where the functions $\Pi_i(E,m,\theta)$ and $\Gamma_j(E,\beta)$ ($i=1,2,3,4$ and $j=1,2,3$) were explicitly written in Appendix \eqref{appendicea} for improved presentation of the results. From Figure \eqref{graf1}, it is evident that the thermal corrections to the differential cross-section are of critical importance at very high temperatures (as $\beta$ approaches zero), where $\Gamma_2$ and $\Gamma_3$ are the dominant factors. It can be shown that in this regime, $(d\sigma/d\Omega)_{\beta}$ increases proportionally to the square of the temperature $T$. In the zero-temperature limit, on the other hand, $\Gamma_2$ and $\Gamma_3$ tend to zero, while $\Gamma_1$ tends to one. 

\begin{figure}[t!]
    \centering
    \begin{minipage}{0.47\textwidth}
        \includegraphics[width=\linewidth]{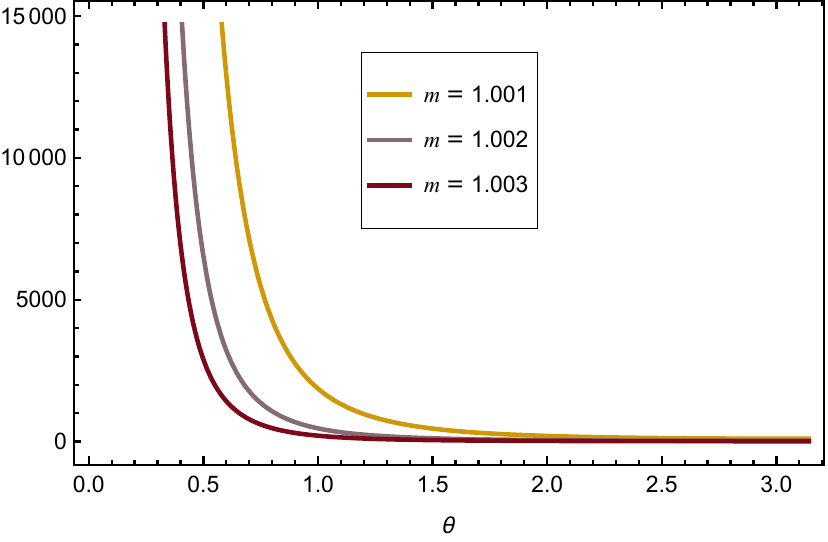}
        \captionsetup{width=\linewidth}
        \caption{\justifying Angular dependence of the differential cross-section at zero temperature for $E\approx m$. We have used $E=1$ and $e=1$.}
        \label{graf3}
    \end{minipage}
    \hfill
    \begin{minipage}{0.47\textwidth}
        \includegraphics[width=\linewidth]{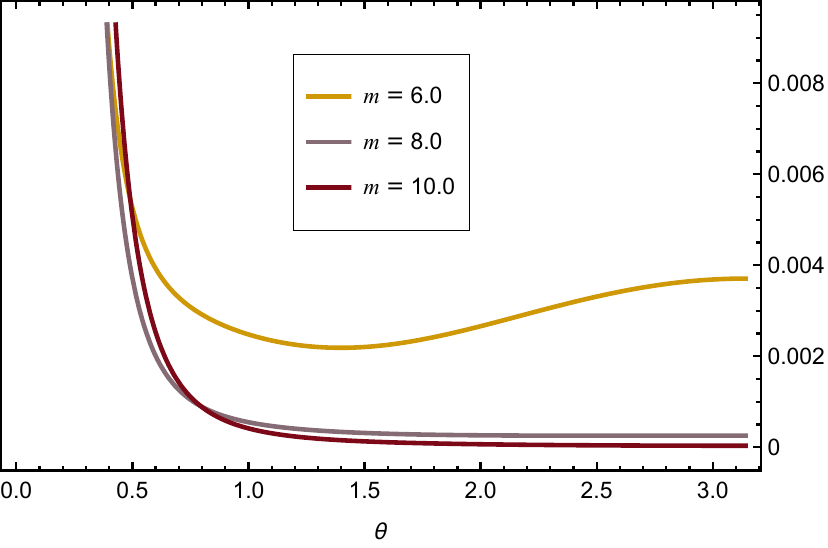}
        \captionsetup{width=\linewidth}
        \caption{\justifying Angular dependence of the differential cross-section at zero temperature for $E>m$. We have used $E=14$ and $e=1$.}
        \label{graf4}
    \end{minipage}
\end{figure}

In Figures \eqref{graf2}, \eqref{graf3}, and \eqref{graf4}, the differential cross section at zero temperature is depicted as a function of $\theta$ for several values of the mass $m$ in three distinct regimes: $E<m$, $E\approx m$, and $E>m$, respectively. Although the three cases exhibit completely different behaviors, some general features deserve mention. First, note that the particle mass directly influences the value assumed by $(d\sigma/d\Omega)_{\beta \rightarrow \infty}$. When $E<m$, for example, it is observed that the larger the value of $m$, the greater $(d\sigma/d\Omega)_{\beta \rightarrow \infty}$ will be (at least for values of $\theta$ that are not too small), although it assumes a minimum near $\theta = \pi /2$. For $E \approx m$, we observe a behavior opposite to that shown in the previous case, meaning that the smaller the value of $m$,  the larger $(d\sigma/d\Omega)_{\beta \rightarrow \infty}$ becomes, although it tends to approach relatively similar values in the limit as $\theta$ approaches $\pi$. In the case where $E > m$, for values of $\theta$ that are not too small, we observe that the zero temperature differential cross-section increases for smaller values of the mass $m$, exhibiting a behavior opposite to that shown in Figure \eqref{graf2}. It is also important to highlight that in all three regimes analyzed so far, for considerably small values of $\theta$, the differential cross-section at zero temperature will be larger for values of $m$ increasingly close to $E$. This is due to the fact that in this limit, $(d\sigma/d\Omega)_{\beta \rightarrow \infty}$ scales proportionally to $(E^2-m^2)^{-2}\, \theta^4$. 

\begin{figure}[t!]
    \centering
    \begin{minipage}{0.47\textwidth}
        \includegraphics[width=\linewidth]{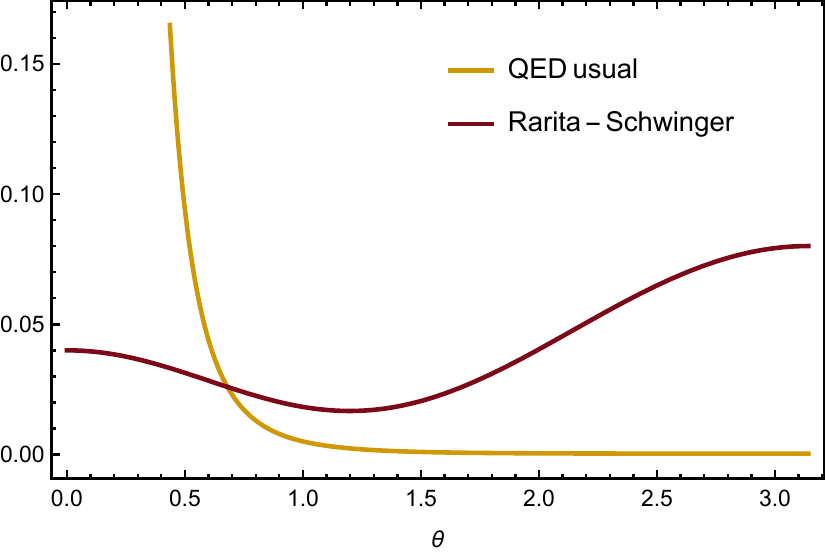}
        \captionsetup{width=\linewidth}
        \caption{\justifying Angular dependence of the differential cross section at zero temperature for usual QED (yellow line) and for the Rarita-Schwinger model (red line). We have used $E=2$ and $e=m=1$.}
        \label{graf5}
    \end{minipage}
    \hfill
    \begin{minipage}{0.47\textwidth}
        \includegraphics[width=\linewidth]{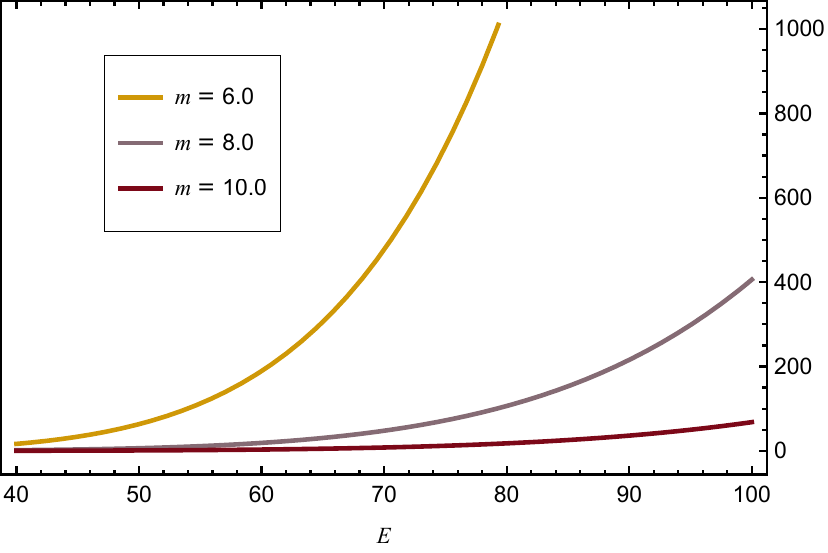}
        \captionsetup{width=\linewidth}
        \caption{\justifying Energy dependence of the total cross section in the ultrarelativistic limit at zero temperature for several values of the mass $m$ in the Rarita-Schwinger model. We have used $e=1$ for this plot.}
        \label{graf6}
    \end{minipage}
\end{figure}

In Figure \eqref{graf5}, we plot the differential cross-sections in the ultra-relativistic limit at zero temperature for the usual QED (involving spin-$1/2$ fermions) and for the Rarita-Schwinger model (involving spin-$3/2$ fermions) considered here. As we can see, the result corresponding to the Rarita-Schwinger model is well-behaved for all values of $\theta$, unlike the one corresponding to usual QED, which diverges for extremely small values of $\theta$. For large values of $\theta$, it is observed that the Rarita-Schwinger model dominates, while for small $\theta$ the behavior reverses. It is also important to highlight that the good behavior of the Rarita-Schwinger model allows us to calculate the total cross-section at zero temperature in the ultra-relativistic limit as 
\begin{eqnarray}
    \sigma_{T=0}(E \gg m) = \frac{26\,  e^4\,  E^6}{1215\,  \pi \,  m^8} .
\end{eqnarray} 
In Figure \eqref{graf6}, we plot the result above as a function of the energy $E$ for several values of the mass $m$. As we can see, the smaller the value of the mass, the larger $ \sigma_{T=0}$ becomes. 

\begin{figure}[!]
     \centering
     \includegraphics[width=0.5\linewidth]{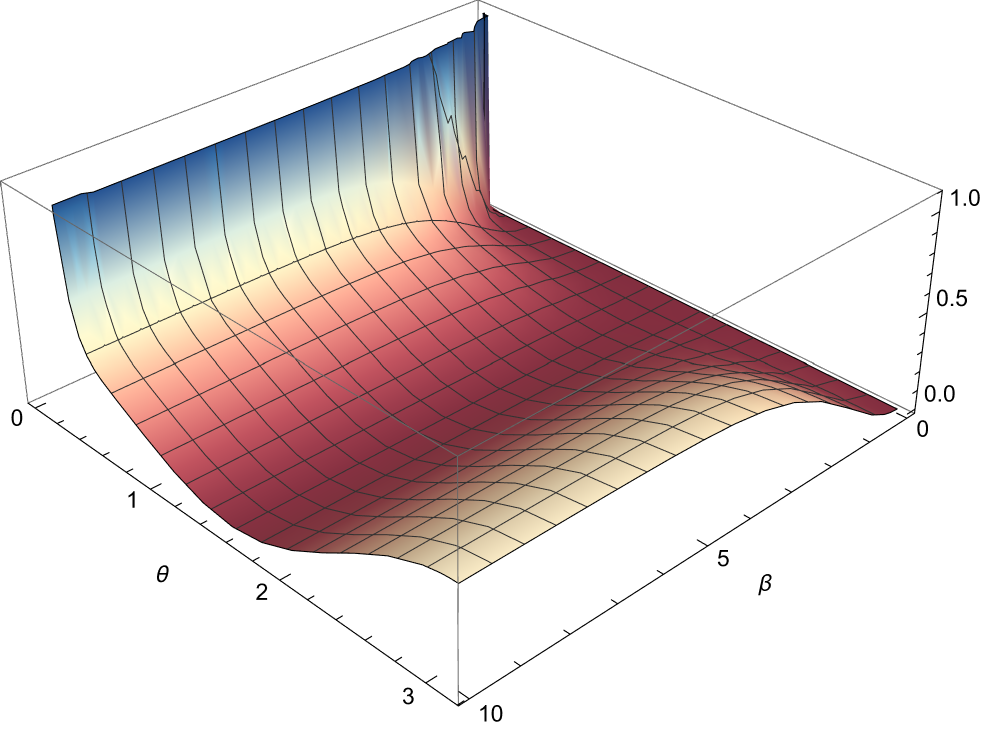}
        \captionsetup{width=\linewidth}
        \caption{\justifying Dependence on $\beta$ and $\theta$ of the well-behaved sector of the differential cross section at finite temperature in the Rarita-Schwinger model. We have used $E=1$, $m=6$, and $e=1$ for this plot.}
        \label{graf7}
\end{figure}

Returning to Eq. \eqref{secdiftermalfinal1111}, we observe that in the result of the temperature-dependent differential cross-section, there exists a product of Dirac delta functions with the same argument. Such terms are characteristic of the TFD formalism and represent a type of apparent singularity, which can be appropriately addressed through a regularization scheme involving derivatives of the delta function, i.e., 
\begin{eqnarray}\label{propriedadedaderivadadadeltaprop}
 2\pi i \frac{1}{n!}\frac{\partial^n}{\partial x^n }\delta (x) = \left( -\frac{1}{x+i\epsilon} \right)^{n+1} - \left( -\frac{1}{x-i\epsilon} \right)^{n+1}.
\end{eqnarray}
Finally, we plot in Figure \eqref{graf7} the $\beta$ and $\theta$-dependencies of the well-behaved part (terms not involving deltas) of the result in Eq. \eqref{secdiftermalfinal1111}. As can be seen, this contribution is completely suppressed in the limit of extremely high temperatures ($\beta \rightarrow 0$), as expected from Figure \eqref{graf1}.

% \begin{figure}[!]
%      \centering
%      \begin{subfigure}[b]{0.49\textwidth}
%          \centering
%          \includegraphics[height=5cm]{graf1.pdf}
%          \caption{}
%          \label{graf1}
%      \end{subfigure}
%     \hfill
%      \begin{subfigure}[b]{0.49\textwidth}
%          \centering
%          \includegraphics[height=4.95cm]{graf2.pdf}
%          \caption{}
%          \label{graf2}
%      \end{subfigure}
%      \hfill
%      \begin{subfigure}[b]{0.49\textwidth}
%          \centering
%          \includegraphics[height=4.95cm]{graf3.pdf}
%          \caption{}
%          \label{graf3}
%      \end{subfigure}
%      \hfill
%      \begin{subfigure}[b]{0.49\textwidth}
%          \centering
%          \includegraphics[height=4.95cm]{graf4.pdf}
%          \caption{}
%          \label{graf4}
%      \end{subfigure}
%      \hfill
%      \begin{subfigure}[b]{0.49\textwidth}
%          \centering
%          \includegraphics[height=4.95cm]{graf5.pdf}
%          \caption{}
%          \label{graf5}
%      \end{subfigure}
%      \hfill
%      \begin{subfigure}[b]{0.49\textwidth}
%          \centering
%          \includegraphics[height=4.95cm]{graf6.pdf}
%          \caption{}
%          \label{graf6}
%      \end{subfigure}
% \end{figure}

%-----$--------------------------------------------------------------------
\section{Conclusion}\label{conclusion}
 
In this paper, we studied Bhabha scattering in a massive Rarita-Schwinger model at finite temperature. The Bhabha scattering was investigated at tree level, and the finite-temperature effects were addressed by employing the TFD formalism. We considered the usual fermion-antifermion into fermion-antifermion scattering and computed the cross section in order to investigate the influence of the finite temperature effects.

To compute the Bhabha-like differential cross section at finite temperature, we  taken into account the modification of the fermion-photon vertex imposed by the Rarita-Schwinger model and the proper thermal contribution to the photon propagator in order to allow the TFD formalism to be properly employed. From the computed cross section, it became evident that the thermal corrections to the differential cross section are of critical importance at very high temperatures (as $\beta$ approaches zero). In this regime, $(d\sigma/d\Omega)_{\beta}$ was shown to increase proportionally to the square of the temperature $T$. 

We also investigated the zero temperature differential cross section for the Bhabha-like scaterring. We noticed that the differential cross section at zero temperature exhibits different behaviors in three distinct regimes: $E<m$, $E\approx m$, and $E>m$. The three cases exhibit entirely different behaviors, however some general features are worth mentioning. We have shown that the mass of the particles directly influences the value assumed by $(d\sigma/d\Omega)_{\beta \rightarrow \infty}$. When $E<m$, for example, it is observed that the larger the value of $m$, the greater $(d\sigma/d\Omega)_{\beta \rightarrow \infty}$ is (at least for values of $\theta$ that are not too small). And for $E \approx m$, we observed a behavior in which the smaller the value of $m$, the larger $(d\sigma/d\Omega)_{\beta \rightarrow \infty}$ becomes, although it tends to approach relatively similar values in the limit as $\theta$ approaches $\pi$. In the case where $E > m$, for values of $\theta$ that are not too small, we observed that the zero-temperature differential cross section increases for smaller values of the mass $m$.

Finally, we studied the differential cross-sections in the ultra-relativistic limit at zero temperature for the usual QED (involving spin-$1/2$ fermions) and for the Rarita-Schwinger model (involving spin-$3/2$ fermions) considered here. We found that the result corresponding to the Rarita-Schwinger model is well-behaved for all values of $\theta$, unlike the one corresponding to the usual QED, which diverges for extremely small values of $\theta$. For large values of $\theta$, it was observed that the Rarita-Schwinger model dominates, while for small $\theta$ the behavior reverses.

\section*{Acknowledgments}
\hspace{0.5cm} JGL would like to thank the Coordena\c{c}\~{a}o de Aperfei\c{c}oamento de Pessoal de N\'{i}vel Superior (CAPES) - Finance Code 001. JF would like to thank the Funda\c{c}\~{a}o Cearense de Apoio ao Desenvolvimento Cient\'{i}fico e Tecnol\'{o}gico (FUNCAP) under the grant PRONEM PNE0112-00085.01.00/16 and the National Council for scientific and technological support (CNPq) under the grant 304485/2023-3 for financial support. TM would like to thank the FAPEAL project No. E:60030.0000002341/2022 and CNPq project No. 316499/2021-8 for financial support.

\vspace{0.5cm}

\center{\bf No Data associated in the manuscript}

\appendix

\section{Explicit form of the functions $\Pi_i(E,m,\theta)$ and $\Gamma_j(E,m,\theta)$}\label{appendicea}

\begin{eqnarray}
  \Pi_1(E,m,\theta) &=& -\frac{1}{(E^2-m^2)^2}\Bigg[ 
-2 E^4 \left(E^2-m^2\right)^6 \cos (4 \theta )-162 E^4 m^8 \left(m^2-2 E^2\right)^2 \csc ^4\left(\frac{\theta
   }{2}\right)\nonumber\\
   &-&\left(E^2-m^2\right)^5 \left(4 E^3+3 E m^2\right)^2 \cos (3 \theta )\nonumber\\
   &+& E^2
   \left(E^2-m^2\right)^3 \left(144 E^8-520 E^6 m^2-871 E^4 m^4-34 E^2 m^6-9 m^8\right) \cos (\theta
   )\nonumber\\
   &-& \left(E^2-m^2\right)^4 \left(120 E^8-160 E^6 m^2+218 E^4 m^4-126 E^2 m^6+81 m^8\right) \cos (2
   \theta )\nonumber\\
   &+& 9 E^2 m^4 (E^2-m^2) \left(32 E^8-160 E^6 m^2+204 E^4 m^4-28 E^2 m^6-9
   m^8\right) \csc ^2\left(\frac{\theta }{2}\right)\nonumber\\
   &-&\left(E^2-m^2\right)^2 \left(262 E^{12}+24 E^{10} m^2+338
   E^8 m^4-1102 E^6 m^6\right. \nonumber\\
   &-& \left. 3 E^4 m^8+54 E^2 m^{10}+81 m^{12}\right) \Bigg]\nonumber\\
   \\
   \nonumber\\
   \Pi_2(E,m,\theta) &=& 20 E^{12}-24 E^{10} m^2+136 E^8 m^4-120 E^6 m^6+265 E^4 m^8-174 E^2 m^{10}\nonumber\\
    &+& 4
   \left(E^2-m^2\right)^6 \cos ^6(\theta )-8 \left(E^2-m^2\right)^6 \cos ^5(\theta )\nonumber\\
   &+& 4 \left(E^2-m^2\right)^4 \left(3
   E^4-14 E^2 m^2+6 m^4\right) \cos ^4(\theta )\nonumber\\
   &-& 8 \left(E^2-m^2\right)^3 \left(6 E^6-18 E^4 m^2-2
   E^2 m^4-m^6\right) \cos ^3(\theta )\nonumber\\
   &+& \left(E^2-m^2\right)^2 \left(92 E^8-176 E^6 m^2+32 E^4 m^4-136
   E^2 m^6+69 m^8\right) \cos ^2(\theta )\nonumber\\
   &-& 2 \left(36 E^{12}-88 E^{10} m^2+156 E^8 m^4-228 E^6 m^6 \right. \nonumber\\
   &+& \left. 95
   E^4 m^8+40 E^2 m^{10}-11 m^{12}\right) \cos (\theta )+59 m^{12}
   \nonumber\\
   \\
   \nonumber\\
   \Pi_3(E,m,\theta)  &=& 2 \left(E^2-m^2\right)^4 \left(4 E^2+3 m^2\right)^2 \cos ^4(\theta )\nonumber\\
  &+& 4 m^2 \left(E^2-m^2\right)^3 \left(16
   E^4-2 E^2 m^2+9 m^4\right) \cos ^3(\theta )\nonumber\\
   &-& \left(E^2-m^2\right)^2 \left(64 E^8-160 E^6 m^2-284
   E^4 m^4-16 E^2 m^6-9 m^8\right) \cos ^2(\theta )\nonumber\\
   &-& 2 m^2 \left(32 E^{10}+84 E^8 m^2-314 E^6 m^4+289
   E^4 m^6-127 E^2 m^8+36 m^{10}\right) \cos (\theta )\nonumber\\
   &+& E^2 \left(32 E^{10}-208 E^8 m^2+738
   E^6 m^4-864 E^4 m^6+601 E^2 m^8-164 m^{10}\right)\nonumber\\
\end{eqnarray}

\begin{eqnarray}
    \Pi_4(E,m,\theta)  &=& 64 E^{12}-176 E^8 m^4+336 E^6 m^6-239 E^4 m^8+90 E^2 m^{10}\nonumber\\
    &+& \left(8 E^6-16 E^4 m^2+17
   E^2 m^4-9 m^6\right)^2 \cos ^2(\theta )\\
   \nonumber\\
    \Gamma_1(E,\beta) &=& \tanh ^2\left(\frac{\beta  \text{En}}{2}\right)\\
    \nonumber\\
    \Gamma_2(E,\beta) &=&  \tanh ^2\left(\frac{\beta  \text{En}}{2}\right) (\coth (\beta  \text{En})-1)^2 \coth ^2(\beta  \text{En})\\
    \nonumber\\
     \Gamma_3(E,\beta) &=& \frac{\cosh ^2(\beta  \text{En}) \tanh ^2\left(\frac{\beta  \text{En}}{2}\right) (\sinh (2 \beta  \text{En})+\cosh (2 \beta 
   \text{En}))}{(\sinh (2 \beta  \text{En})+\cosh (2 \beta  \text{En})-1)^4}
\end{eqnarray}

\global\long\def\link#1#2{\href{http://eudml.org/#1}{#2}}
 \global\long\def\doi#1#2{\href{http://dx.doi.org/#1}{#2}}
 \global\long\def\arXiv#1#2{\href{http://arxiv.org/abs/#1}{arXiv:#1 [#2]}}
 \global\long\def\arXivOld#1{\href{http://arxiv.org/abs/#1}{arXiv:#1}}

%%%%%%%%%%%%%%%%%%%%%%%%%%%%%%%%%%%%%%%%%%%%%%%%%%%%%%%%%%%%%%%%%%%%%%%%%%%%


\begin{thebibliography}{99}

\bibitem{Bengtsson:2020}
A. Bengtsson, ``Higher Spin Field Theory'',
\doi{10.1515/9783110451771}{Texts and Monographs in Theoretical Physics, De
Gruyter (2020)}.

%\bibitem{Bonora:2016ida}
%L.~Bonora, M.~Cvitan, P.~Dominis Prester, B.~Lima de Souza and I.~Smoli\'c,
%``Massive fermion model in 3d and higher spin currents,''
%\doi{10.1007/JHEP05(2016)072} {JHEP \textbf{05} (2016), 072}
%\arXiv{1602.07178}{hep-th}.

\bibitem{Bonora:2016otz}
L.~Bonora, M.~Cvitan, P.~Dominis Prester, S.~Giaccari, B.~Lima de Souza and T.~\v{S}temberga,
``One-loop effective actions and higher spins,''
\doi{10.1007/JHEP12(2016)084}{JHEP \textbf{12} (2016), 084}
[\arXiv{1609.02088}{hep-th}].

\bibitem{Bonora:2017ykb}
L.~Bonora, M.~Cvitan, P.~Dominis Prester, S.~Giaccari and T.~\v{S}temberga,
``One-loop effective actions and higher spins. Part II,''
\doi{10.1007/JHEP01(2018)080}{JHEP \textbf{01} (2018), 080}
[\arXiv{1709.01738}{hep-th}].

\bibitem{Rahman:2012thy}
R.~Rahman,
``Higher Spin Theory - Part I,''
\doi{10.22323/1.195.0004}{PoS \textbf{ModaveVIII}, 004 (2012)}
[\arXiv{1307.3199}{hep-th}].

\bibitem{Adler:2014pga}
S.~L.~Adler,
``SU(8) family unification with boson-fermion balance,''
\doi{10.1142/S0217751X14501309}{Int. J. Mod. Phys. A \textbf{29} (2014) 1450130}
[\arXiv{1403.2099}{hep-th}].

\bibitem{Campoleoni:2024ced}
A.~Campoleoni and S.~Fredenhagen,
``Higher-spin gauge theories in three spacetime dimensions''
[\arXiv{2403.16567}{hep-th}].

\bibitem{Giombi:2016ejx}
S.~Giombi,
``Higher Spin \textemdash{} CFT Duality''
[\arXiv{1607.02967}{hep-th}].

\bibitem{Singh:1974qz}
L.~P.~S.~Singh and C.~R.~Hagen,
``Lagrangian formulation for arbitrary spin. 1. The boson case,''
\doi{10.1103/PhysRevD.9.898}{Phys. Rev. D \textbf{9} (1974), 898-909}.

\bibitem{Singh:1974rc}
L.~P.~S.~Singh and C.~R.~Hagen,
``Lagrangian formulation for arbitrary spin. 2. The fermion case,''
\doi{10.1103/PhysRevD.9.910}{Phys. Rev. D \textbf{9} (1974), 910-920}.

\bibitem{Fronsdal:1978rb}
C.~Fronsdal,
``Massless Fields with Integer Spin,''
\doi{10.1103/PhysRevD.18.3624}{Phys. Rev. D \textbf{18} (1978), 3624}.

\bibitem{Fang:1978wz}
J.~Fang and C.~Fronsdal,
``Massless Fields with Half Integral Spin,''
\doi{10.1103/PhysRevD.18.3630}{Phys. Rev. D \textbf{18} (1978), 3630}.

\bibitem{Rarita:1941mf}
W.~Rarita and J.~Schwinger,
``On a theory of particles with half integral spin,''
\doi{10.1103/PhysRev.60.61}{Phys. Rev. \textbf{60} (1941), 61}.

\bibitem{Freedman:1976py}
D.~Z.~Freedman and P.~van Nieuwenhuizen,
``Properties of Supergravity Theory,''
\doi{10.1103/PhysRevD.14.912}{Phys. Rev. D \textbf{14} (1976), 912}.

\bibitem{Das:1976ct}
A.~K.~Das and D.~Z.~Freedman,
``Gauge Quantization for Spin 3/2 Fields,''
\doi{10.1016/0550-3213(76)90589-7}{Nucl. Phys. B \textbf{114} (1976), 271-296}.

\bibitem{Gates:1983nr}
S.~J.~Gates, M.~T.~Grisaru, M.~Rocek and W.~Siegel,
``Superspace Or One Thousand and One Lessons in Supersymmetry,''
Front. Phys. \textbf{58} (1983), Addison-Wesley
[\arXivOld{hep-th/0108200}].

\bibitem{Delgado-Acosta:2009ulg}
E.~G.~Delgado-Acosta and M.~Napsuciale,
``Compton scattering off elementary spin 3/2 particles,''
\doi{10.1103/PhysRevD.80.054002}{Phys. Rev. D \textbf{80} (2009), 054002}
[\arXiv{0907.1124}{hep-th}].

\bibitem{Antoniadis:2022jjy}
I.~Antoniadis, A.~Guillen and F.~Rondeau,
``Massive gravitino scattering amplitudes and the unitarity cutoff of the new Fayet-Iliopoulos terms,''
\doi{10.1007/JHEP01(2023)043}{JHEP \textbf{01}, 043 (2023)}
[\arXiv{2210.00817}{hep-th}].

\bibitem{deJong:1992wm}
F.~de Jong and R.~Malfliet,
``Covariant description of the Delta in nuclear matter,''
\doi{10.1103/PhysRevC.46.2567}{Phys. Rev. C \textbf{46} (1992), 2567-2581}.

\bibitem{Pascalutsa:1999zz}
V.~Pascalutsa and R.~Timmermans,
``Field theory of nucleon to higher spin baryon transitions,''
\doi{10.1103/PhysRevC.60.042201}{Phys. Rev. C \textbf{60} (1999), 042201}
[\arXivOld{nucl-th/9905065}].

\bibitem{Bernard:2003xf}
V.~Bernard, T.~R.~Hemmert and U.~G.~Meissner,
``Infrared regularization with spin 3/2 fields,''
\doi{10.1016/S0370-2693(03)00538-0}{Phys. Lett. B \textbf{565} (2003), 137-145}
[\arXivOld{hep-ph/0303198}].

\bibitem{Gomes:2022btc}
M.~Gomes, T.~Mariz, J.~R.~Nascimento and A.~Y.~Petrov,
``Lorentz-breaking Rarita-Schwinger model,''
\doi{10.1088/1402-4896/ad0e9c}{Phys. Scripta \textbf{98} (2023) no.12, 125260}
[\arXiv{2211.12414}{hep-th}].

\bibitem{Gomes:2023qkj}
M.~Gomes, J.~G.~Lima, T.~Mariz, J.~R.~Nascimento and A.~Y.~Petrov,
``Non-Abelian Carroll\textendash{}Field\textendash{}Jackiw term in a Rarita-Schwinger model,''
\doi{10.1016/j.physletb.2023.138141}{Phys. Lett. B \textbf{845} (2023), 138141}
[\arXiv{2308.16308}{hep-th}].

\bibitem{Moldauer:1956zz}
P.~A.~Moldauer and K.~M.~Case,
``Properties of Half-Integral Spin Dirac-Fierz-Pauli Particles,''
\doi{10.1103/PhysRev.102.279}{Phys. Rev. \textbf{102} (1956), 279-285}.

\bibitem{Johnson:1960vt}
K.~Johnson and E.~C.~G.~Sudarshan,
``Inconsistency of the local field theory of charged spin 3/2 particles,''
\doi{10.1016/0003-4916(61)90030-6}{Annals Phys. \textbf{13} (1961), 126-145}.

\bibitem{Velo:1969txo}
G.~Velo and D.~Zwanziger,
``Noncausality and other defects of interaction lagrangians for particles with spin one and higher,''
\doi{10.1103/PhysRev.188.2218}{Phys. Rev. \textbf{188} (1969), 2218-2222}.

\bibitem{Velo:1969bt}
G.~Velo and D.~Zwanziger,
``Propagation and quantization of Rarita-Schwinger waves in an external electromagnetic potential,''
\doi{10.1103/PhysRev.186.1337}{Phys. Rev. \textbf{186} (1969), 1337-1341}.

\bibitem{Aurilia:1969bg}
A.~Aurilia and H.~Umezawa,
``Theory of high-spin fields,''
\doi{10.1103/PhysRev.182.1682}{Phys. Rev. \textbf{182} (1969), 1682-1694}.

\bibitem{Nath:1971wp}
L.~M.~Nath, B.~Etemadi and J.~D.~Kimel,
``Uniqueness of the interaction involving spin 3/2 particles,''
\doi{110.1103/PhysRevD.3.2153}{Phys. Rev. D \textbf{3} (1971), 2153-2161}.

\bibitem{Adler:2015yha}
S.~L.~Adler,
``Classical Gauged Massless Rarita-Schwinger Fields,''
\doi{10.1103/PhysRevD.92.085022}{Phys. Rev. D \textbf{92} (2015) no.8, 085022}
[\arXiv{1508.03380}{hep-th}].

\bibitem{Adler:2015zha}
S.~L.~Adler,
``Quantized Gauged Massless Rarita-Schwinger Fields,''
\doi{10.1103/PhysRevD.92.085023}{Phys. Rev. D \textbf{92} (2015) no.8, 085023}
[\arXiv{1508.03382}{hep-th}].

\bibitem{LeiteLopes:1980mh}
J.~Leite Lopes, J.~A.~Martins Simoes and D.~Spehler,
``Production and Decay Properties of Possible Spin 3/2 Leptons,''
\doi{10.1016/0370-2693(80)90898-9}{Phys. Lett. B \textbf{94}, 367-372 (1980)}.

\bibitem{LeiteLopes:1980pa}
J.~Leite Lopes, J.~A.~Martins Simoes and D.~Spehler,
``Possible Spin 3/2 Quarks and Scaling Violations in Neutrino Reactions,''
\doi{10.1103/PhysRevD.23.797}{Phys. Rev. D \textbf{23}, 797 (1981)}.

\bibitem{LeiteLopes:1981ys}
J.~Leite Lopes, D.~Spehler and J.~A.~Martins Simoes,
``WEAK INTERACTIONS INVOLVING SPIN 3/2 LEPTONS,''
\doi{10.1103/PhysRevD.25.1854}{Phys. Rev. D \textbf{25}, 1854 (1982)}.

\bibitem{Tosa:1985nn}
Y.~Tosa and R.~E.~Marshak,
``EXOTIC FERMIONS,''
\doi{10.1103/PhysRevD.32.774}{Phys. Rev. D \textbf{32}, 774 (1985)}.

%\bibitem {Freedman76}D. Z. Freedman, P. van Nieuwenhuizen and S. Ferrara, Phys. Rev. D \textbf{13} (1976) 3214

\bibitem{Choudhury:1984bu}
S.~R.~Choudhury, R.~G.~Ellis and G.~C.~Joshi,
``LIMITS ON EXCITED SPIN 3/2 LEPTONS,''
\doi{10.1103/PhysRevD.31.2390}{Phys. Rev. D \textbf{31}, 2390 (1985)}.

\bibitem{Almeida:1995yp}
F.~M.~L.~Almeida, Jr., J.~H.~Lopes, J.~A.~Martins Simoes and A.~J.~Ramalho,
``Production and decay of single heavy spin 3/2 leptons in high-energy electron - positron collisions,''
\doi{10.1103/PhysRevD.53.3555}{Phys. Rev. D \textbf{53}, 3555-3558 (1996)}
[\arXiv{9509364}{hep-ph}].

\bibitem{Criado:2021itq}
J.~C.~Criado, A.~Djouadi, N.~Koivunen, M.~Raidal and H.~Veerm\"ae,
``Higher-spin particles at high-energy colliders,''
\doi{10.1007/JHEP05(2021)254}{JHEP \textbf{05}, 254 (2021)}
[\arXiv{2102.13652}{hep-ph}].

\bibitem{Ozdem:2024rqx}
U.~\"Ozdem,
``Elucidating the nature of hidden-charm pentaquark states with spin-32 through their electromagnetic form factors,''
\doi{10.1016/j.physletb.2024.138551}{Phys. Lett. B \textbf{851}, 138551 (2024)}
[\arXiv{2402.03802}{hep-ph}].

\bibitem{Cakir:2007wn}
O.~Cakir and A.~Ozansoy,
``Search for excited spin-3/2 and spin-1/2 leptons at linear colliders,''
\doi{10.1103/PhysRevD.77.035002}{Phys. Rev. D \textbf{77}, 035002 (2008)},
[\arXiv{0709.2134}{hep-ph}].

\bibitem{Ozansoy:2016ivj}
A.~Ozansoy, V.~Ar\i{} and V.~\c{C}etinkaya,
``Search for excited spin-3/2 neutrinos at LHeC,''
\doi{10.1155/2016/1739027}{Adv. High Energy Phys. \textbf{2016}, 1739027 (2016)}
[\arXiv{1607.04437}{hep-ph}].

\bibitem{Araujo:2022qke}
M.~C.~Ara\'ujo and R.~V.~Maluf,
``Meson scattering in a Lorentz-violating scalar QED at finite temperature,''
\doi{10.1016/j.aop.2022.169036}{Annals Phys. \textbf{444} (2022), 169036}
[\arXiv{2205.1451}{hep-ph}].

\bibitem{Ulhoa:2023opw}
S.~C.~Ulhoa, A.~F.~d.~Santos, E.~P.~Spaniol and F.~C.~Khanna,
``On Thermodynamics of Kerr Black Hole,''
\doi{10.1002/andp.202300173}{Annalen Phys. \textbf{535} (2023) no.8, 2300173}
[\arXiv{2306.02950}{gr-qc}].

\bibitem{Xu:2020wky}
Z.~Xu, A.~Chenu, T.~Prosen and A.~del Campo,
``Thermofield dynamics: Quantum Chaos versus Decoherence,''
Phys. Rev. B \textbf{103}, no.6, 064309 (2021)
doi:10.1103/PhysRevB.103.064309
[arXiv:2008.06444 [quant-ph]].

\bibitem{Takahashi:1996zn}
Y.~Takahashi and H.~Umezawa,
``Thermo field dynamics,''
\doi{10.1142/S0217979296000817}{Int. J. Mod. Phys. B \textbf{10}, 1755-1805 (1996)}.

\bibitem{Umezawa:1982nv}
H.~Umezawa, H.~Matsumoto and M.~Tachiki,
``THERMO FIELD DYNAMICS AND CONDENSED STATES,'' (1982).

\bibitem{Araujo:2023izx}
M.~C.~Ara\'ujo, J.~Furtado and R.~V.~Maluf,
``Lorentz-violating extension of scalar QED at finite temperature,''
\doi{10.1016/j.physletb.2023.138064}{Phys. Lett. B \textbf{844} (2023), 138064}
[\arXiv{2306.06959}{hep-th}].

\bibitem{Assuncao:2020cdo}
J.~F.~Assun\c{c}\~ao, J.~Furtado and T.~Mariz,
``Nonanalyticity of the non-Abelian five-dimensional Chern-Simons term,''
\doi{10.1209/0295-5075/134/41002}{EPL \textbf{134} (2021) no.4, 41002}
[\arXiv{2011.12333}{hep-th}].

\bibitem{Cabral:2024tqa}
D.~S.~Cabral and A.~F.~Santos,
``$e^{+}e^{-}\rightarrow l^{+}l^{-}$ scattering at finite temperature in the presence of a classical background magnetic field,''
\doi{10.1140/epjp/s13360-024-04975-w}{Eur. Phys. J. Plus \textbf{139} (2024) no.2, 190}
[\arXiv{2402.01457}{hep-ph}].

\bibitem{Santos:2021hed}
A.~F.~Santos and F.~C.~Khanna,
``Temperature effects for $e^-+e^+\rightarrow \mu ^-+\mu ^+$ scattering in very special relativity,''
\doi{10.1140/epjp/s13360-021-02312-z}{Eur. Phys. J. Plus \textbf{137} (2022) no.1, 101}
[\arXiv{2112.11422}{hep-th}].

\bibitem{Souza:2021uuf}
P.~R.~A.~Souza, A.~F.~Santos and F.~C.~Khanna,
``Effects of the CPT-even and Lorentz violation on the Bhabha scattering at finite temperature,''
\doi{10.1016/j.aop.2021.168451}{Annals Phys. \textbf{428} (2021), 168451}
[\arXiv{2103.08404}{hep-th}].

\bibitem{Araujo:2024hnu}
M.~C.~Ara\'ujo, R.~V.~Maluf and J.~Furtado,
``Meson scattering in a non-minimally Lorentz-violating scalar QED at finite temperature,''
\doi{10.1140/epjc/s10052-024-13235-1}{Eur. Phys. J. C \textbf{84}, no.8, 844 (2024)}
[\arXiv{2406.17078}{hep-th}].

\bibitem{Santos:2022zfz}
A.~F.~Santos and F.~C.~Khanna,
``Non-Abelian Aether-Like Term and Applications at Finite Temperature,''
\doi{10.1155/2022/6703645}{Adv. High Energy Phys. \textbf{2022} (2022), 6703645}
[\arXiv{2209.09727}{hep-th}].

\bibitem{Pilling:2004cu}
T.~Pilling,
``Symmetry of massive Rarita-Schwinger fields,''
\doi{10.1142/S0217751X05021300}{Int. J. Mod. Phys. A \textbf{20} (2005), 2715-2742}
[\arXivOld{hep-th/0404131}].

\bibitem{Mariz:2010fm}
T.~Mariz,
``Radiatively induced Lorentz-violating operator of mass dimension five in QED,''
\doi{10.1103/PhysRevD.83.045018}{Phys. Rev. D \textbf{83}, 045018 (2011)}
[\arXiv{1010.5013}{hep-th}].

\bibitem{Rubtsov:2012kb}
G.~Rubtsov, P.~Satunin and S.~Sibiryakov,
``On calculation of cross sections in Lorentz violating theories,''
\doi{10.1103/PhysRevD.86.085012}{Phys. Rev. D \textbf{86}, 085012 (2012)}
[\arXiv{1204.5782}{hep-ph}].

\bibitem{Kostelecky:2009zp}
V.~A.~Kostelecky and M.~Mewes,
``Electrodynamics with Lorentz-violating operators of arbitrary dimension,''
\doi{10.1103/PhysRevD.80.015020}{Phys. Rev. D \textbf{80}, 015020 (2009)}
[\arXiv{0905.0031}{hep-ph}].

\bibitem{Landsman:1986uw}
N.~P.~Landsman and C.~G.~van Weert,
``Real and Imaginary Time Field Theory at Finite Temperature and Density,''
\doi{10.1016/0370-1573(87)90121-9}{Phys. Rept. \textbf{145} (1987), 141}.

\bibitem{van2001introduction}
Ch.~G.~van Weert, 
``An Introduction to Real- and Imaginary-time Thermal Field Theory, Lecture notes on Statistical Field Theory,'' (2001).

%%%%%%%%%%%%%%%%%%%%%%%%%%%%%%%%%%%%%%%%%%%%%%%%%%%%%%%%%%%%%%%%%%%%%%%%%%

\newpage
%\begin{figure}[h!]
%\centering
%\includegraphics[scale=0.6]{difsectime1.pdf}
%\caption{Angular and temperature dependencies of the differential cross section for $a$-type meson-antimeson scattering in $b$-type meson-antimeson in the timelike case. For this plot we have considered $e=1$ and  $E_{cm} = 1$.}\label{difsectime1}
%\end{figure}

% \begin{figure}[h!]
%      \centering
%      \begin{subfigure}[b]{0.3\textwidth}
%          \centering
%          \includegraphics[scale=0.6]{difsecspace1a.pdf}
%          \caption{}
%          \label{difsecspace1a}
%      \end{subfigure}
%      \hfill
%      \begin{subfigure}[b]{0.6\textwidth}
%          \centering
%          \includegraphics[scale=0.6]{difsecspace1b.pdf}
%          \caption{}
%          \label{difsecspace1b}
%      \end{subfigure}
%      \hfill
%      \begin{subfigure}[b]{0.6\textwidth}
%          \centering
%          \includegraphics[scale=0.6]{difsecspace1c.pdf}
%          \caption{}
%          \label{difsecspace1c}
%      \end{subfigure}
%         \caption{Angular dependencies of the zero-temperature differential cross section for $a$-type meson-antimeson scattering in $b$-type meson-antimeson in the spacelike case. For this plot we took $e=1$,  $E_{cm} = 1$. In \eqref{difsecspace1a}, we have the usual case ($g|\mathbf{w}| =0$). In \eqref{difsecspace1b} and \eqref{difsecspace1c}, we have considered $g|\mathbf{w}| =0.9$, but for the former $\theta_{\omega z}=\frac{\pi}{2}$ and $\phi_{\omega x}=0$, while for the latter $\theta_{\omega z}=\phi_{\omega x}=\frac{\pi}{4}$.}
%         \label{fig3space}
% \end{figure}

% \begin{figure}[h!]
%      \centering
%      \begin{subfigure}[b]{0.3\textwidth}
%          \centering
%          \includegraphics[scale=0.6]{difsectime2a.pdf}
%          \caption{}
%          \label{difsectime2a}
%      \end{subfigure}
%      \hfill
%      \begin{subfigure}[b]{0.6\textwidth}
%          \centering
%          \includegraphics[scale=0.6]{difsectime2b.pdf}
%          \caption{}
%          \label{difsectime2b}
%      \end{subfigure}
%         \caption{Angular and temperature dependencies of the differential cross section for $a$-type meson-antimeson scattering in $a$-type meson-antimeson in the timelike case. For this plot we took $e=1$ and  $E_{cm} = 1$. In \eqref{difsectime2a}, no Lorentz violation has been considered ($g \omega_0=0$). In \eqref{difsectime2b}, we have set $g \omega_0=3$. }
%         \label{fig4timecomsemlv}
% \end{figure}

% \begin{figure}[h!]
%      \centering
%      \begin{subfigure}[b]{0.3\textwidth}
%          \centering
%          \includegraphics[scale=0.6]{difsecspace2a.pdf}
%          \caption{}
%          \label{difsecspace2a}
%      \end{subfigure}
%      \hfill
%      \begin{subfigure}[b]{0.6\textwidth}
%          \centering
%          \includegraphics[scale=0.6]{difsecspace2b.pdf}
%          \caption{}
%          \label{difsecspace2b}
%      \end{subfigure}
%      \hfill
%      \begin{subfigure}[b]{0.6\textwidth}
%          \centering
%          \includegraphics[scale=0.6]{difsecspace2c.pdf}
%          \caption{}
%          \label{difsecspace2c}
%      \end{subfigure}
%         \caption{Angular dependencies of the zero-temperature differential cross section for $a$-type meson-antimeson scattering in $a$-type meson-antimeson in the spacelike case. For this plot we took $e=1$,  $E_{cm} = 1$. In \eqref{difsecspace2a}, we have the usual case ($g|\mathbf{w}| =0$). In \eqref{difsecspace2b} and \eqref{difsecspace2c}, we have considered $g|\mathbf{w}| =2$, but for the former $\theta_{\omega z}=\frac{\pi}{2}$ and $\phi_{\omega x}=0$, while for the latter $\theta_{\omega z}=\frac{\pi}{2}$ and $\phi_{\omega x}=\pi$.}
%         \label{fig5difsecspaceaaaabbb}
% \end{figure}

% \begin{figure}[h!]
%      \centering
%      \begin{subfigure}[b]{0.3\textwidth}
%          \centering
%          \includegraphics[scale=0.6]{difsectime3a.pdf}
%          \caption{}
%          \label{difsectime3a}
%      \end{subfigure}
%      \hfill
%      \begin{subfigure}[b]{0.6\textwidth}
%          \centering
%          \includegraphics[scale=0.6]{difsectime3b.pdf}
%          \caption{}
%          \label{difsectime3b}
%      \end{subfigure}
%         \caption{Angular and temperature dependencies of the differential cross section for $a$-type meson-meson scattering in $a$-type meson-meson in the timelike case. For this plot we took $e=1$ and  $E_{cm} = 1$. In \eqref{difsectime3a}, no Lorentz violation has been considered ($g \omega_0=0$). In \eqref{difsectime3b}, we have set $g \omega_0=5$.}
%         \label{fig6timelikediseccccc}
% \end{figure}

% \begin{figure}[h!]
%      \centering
%      \begin{subfigure}[b]{0.3\textwidth}
%          \centering
%          \includegraphics[scale=0.6]{difsecspace3a.pdf}
%          \caption{}
%          \label{difsecspace3a}
%      \end{subfigure}
%      \hfill
%      \begin{subfigure}[b]{0.6\textwidth}
%          \centering
%          \includegraphics[scale=0.6]{difsecspace3b.pdf}
%          \caption{}
%          \label{difsecspace3b}
%      \end{subfigure}
%      \hfill
%      \begin{subfigure}[b]{0.6\textwidth}
%          \centering
%          \includegraphics[scale=0.6]{difsecspace3c.pdf}
%          \caption{}
%          \label{difsecspace3c}
%      \end{subfigure}
%         \caption{Angular dependencies of the zero-temperature differential cross section for $a$-type meson-meson scattering in $a$-type meson-meson in the spacelike case. For this plot we took $e=1$,  $E_{cm} = 1$. In \eqref{difsecspace3a}, we have the usual case ($g|\mathbf{w}| =0$). In \eqref{difsecspace3b} and \eqref{difsecspace3c}, we have considered $g|\mathbf{w}| =5$, but for the former $\theta_{\omega z}=\frac{\pi}{2}$ and $\phi_{\omega x}=0$, while for the latter $\theta_{\omega z}=\frac{\pi}{2}$ and $\phi_{\omega x}=\frac{\pi}{2}$.}
%         \label{fig7difsecspacecccccultcross}
% \end{figure}


\end{thebibliography}
\end{document}